\documentclass[prb,twocolumn,superscriptaddress,floatfix,showpacs,amsmath,amssymb]{revtex4}% Physical Review

\usepackage{graphicx,color}% Include figure files
\usepackage{dcolumn}% Align table columns on decimal point
\newcolumntype{.}{D{.}{.}{-1}}
\usepackage{bm}% bold math
\usepackage[latin1]{inputenc}
\usepackage{epstopdf}
\usepackage{subfigure}

\newcommand{\cp}{$c_{\rm p}$}

\newcommand{\mb}{\(\mu _{\rm B}\)}
\newcommand{\rk}[1]{\textcolor{black}{#1}}

\newcommand{\tn}{$T_{\rm N}$}
\newcommand{\ttwo}{$T_{\rm N2}$}

\newcommand{\zn}{ZnSb$_2$O$_6$}
\newcommand{\mn}{MnSb$_2$O$_6$}
\newcommand{\as}{MnAs$_2$O$_6$}

\newcommand{\cpm}{$c_{\rm p}^{\rm magn}$}
\newcommand{\jmk}{J/(mol$\cdot$K)}

\newcommand{\bsf}{$B_{\rm SF}$}
\newcommand{\bs}{$B_{\rm sat}$}
\newcommand{\be}{$B_{\rm E}$}
\newcommand{\ba}{$B_{\rm A}$}

\begin{document}

%\preprint{APS/123-QED}
%\title{Dynamic Magnetic Properties of MnSb$_2$O$_6$ Studied by Muon-Spin-Relaxation- and High-Frequency Electron-Spin-Resonance Spectroscopy}
\title{Magnetism and the phase diagram of MnSb$_2$O$_6$}

%%%%  AUTHORS %%%%%%%%%%%%%%%%%%%%%%%%%%%%%%%%%%%%%%%%%%%%%%

\author{C.~Koo}
\affiliation{Kirchhoff Institute of Physics, Heidelberg University, D-69120 Heidelberg, Germany}
\author{J.~Werner}
\affiliation{Kirchhoff Institute of Physics, Heidelberg University, D-69120 Heidelberg, Germany}
\author{M.~Tzschoppe}
\affiliation{Kirchhoff Institute of Physics, Heidelberg University, D-69120 Heidelberg, Germany}
\author{ M.~Abdel-Hafiez}
\affiliation{Institute of Physics, Goethe University Frankfurt, 60438 Frankfurt/M, Germany}
\affiliation{National University of Science and Technology "MISiS", Moscow 119049, Russia}
\author{P. K.~ Biswas}
\affiliation{Laboratory for Muon Spin Spectroscopy, Paul Scherrer Institute, CH-5232 Villigen PSI, Switzerland}
\author{R.~Sarkar}
\affiliation{Institute of Solid State and Materials Physics, TU Dresden, D-01069 Dresden, Germany}
%\affiliation{Institute for Solid State Physics, TU Dresden, D-01069 Dresden, Germany.}
\author{H.-H.~Klauss}
\affiliation{Institute of Solid State and Materials Physics, TU Dresden, D-01069 Dresden, Germany}
%\affiliation{Institute for Solid State Physics, TU Dresden, D-01069 Dresden, Germany.}
\author{G. V.~Raganyan}
\affiliation{Physics Faculty, M.V. Lomonosov Moscow State University, Moscow 119991, Russia}
\author{E. A.~Ovchenkov}
\affiliation{Physics Faculty, M.V. Lomonosov Moscow State University, Moscow 119991, Russia}
\author{A. Yu. Nikulin}
\affiliation{Chemistry Faculty, Southern Federal University, 7 ul. Zorge, Rostov-na-Donu, 344090 Russia}
\author{A. N.~Vasiliev}
\affiliation{Physics Faculty, M.V. Lomonosov Moscow State University, Moscow 119991, Russia}
\affiliation{National Research South Ural State University, Chelyabinsk 454080, Russia}
\affiliation{National University of Science and Technology "MISiS", Moscow 119049, Russia}
\author{E. A.~Zvereva}
\affiliation{Physics Faculty, M.V. Lomonosov Moscow State University, Moscow 119991, Russia}
\affiliation{National Research South Ural State University, Chelyabinsk 454080, Russia}
\author{R.~Klingeler}
\email[Email:]{r.klingeler@kip.uni-heidelberg.de}
\affiliation{Kirchhoff Institute of Physics, Heidelberg University, D-69120 Heidelberg, Germany}
\affiliation{Centre for Advanced Materials, Heidelberg University, D-69120 Heidelberg, Germany}

%%%%%%%%%%%%%%%%%%%%%%%%%%%%%%%%%%%%%%%%%%%%%%%%%%

\date{\today}% It is always \today, today,

\begin{abstract}
Static and dynamic magnetic properties of P\={3}1$m$-phase \mn\ have been studied by means of muon spin relaxation ($\mu$SR), high-frequency electron spin resonance (HF-ESR), specific heat, and magnetisation studies in magnetic fields up to 25\,T. The data imply onset of long-range antiferromagnetic order at \tn\ =~8~K and a spin-flop-like transition at \bsf\ $\approx 0.7 - 1$~T. Below \tn, muon asymmetry exhibits well-defined oscillations indicating a narrow distribution of the local fields. A competing antiferromagnetic phase appearing below \ttwo\ =~5.3~K is evidenced by a step in the magnetisation and a slight kink of the relaxation rate. Above \tn , both $\mu$SR and HF-ESR data suggest short-range spin order. HF-ESR data show that local magnetic fields persist up to at least $12\cdot T_{\rm N}\approx 100$\,K. Analysis of the antiferromagnetic resonance modes and the thermodynamic spin-flop field suggest zero-field splitting of $\Delta \approx 18$\,GHz which implies small but finite magnetic anisotropy.
\end{abstract}

\maketitle

\section{Introduction}

Frustrated spin systems provide access to study the emergence of novel ground states and unusual excitations. Triangular Heisenberg antiferromagnets are generic for frustrated magnetism and may possess rich magnetic phase diagrams and unusual ground states.~\cite{Collins,Nakatsuji,Starykh2015} Even in the classical case, i.e., for large spin values, the ground state exhibits a hidden symmetry and unusual excitations.~\cite{Seabra2011} While the tendency towards static magnetic order increases upon partial lifting of frustration, competing interactions still yield unusual ground states and short range spin fluctuations evolve far above \tn . A further ingredient determining the ground states in addition to the spin size and the degree of geometrical frustration is magnetic anisotropy as frustration is more severe when spin alignment is restricted.~\cite{Collins,Yun}

The honeycomb structure in \as\ with P\={3}1$m$ structure exhibits a trigonal arrangement of Mn$^{2+}$-ions and is an example of a frustrated spin system as it exhibits a network of $S=5/2$ triangles with competing antiferromagnetic exchange interactions.~\cite{Wangbo} Indeed, the layered and frustrated magnetic structure gives rise to an incommensurate spin order below 12~K.~\cite{Nakua} The recently synthesized P\={3}1$m$ phase of \mn\ is isostructural and isoelectronic to \as .~\cite{Nalbandyan2015} Note, that this phase differs from the trigonal, structurally and magnetically chiral polymorph of \mn .~\cite{Reimers,Johnson,Werner2017} Density functional theory calculations on the P\={3}1$m$ phase suggest that the exchange interactions are very similar to those of its counterpart \as\ and that
the long range antiferromagnetically ordered ground state appearing below $\sim 8$~K is of incommensurate nature.~\cite{Wangbo} In addition, weak ferromagnetism was suggested from static susceptibility data to be present below $T=41$~K. Here, we report the magnetic phase diagram and a detailed study of the magnetic properties of \mn\ by means of zero-field muon spin resonance ($\mu$SR), high-frequency electron spin resonance (HF-ESR), and static magnetisation measurements. Our data imply long-range antiferromagnetic order of most presumingly commensurate nature. In addition, there is thermodynamic evidence for a competing AFM phase below $T_{\rm N2}=5.3$~K. While weak ferromagnetism is ruled out by our experiments, there is clear evidence for short range magnetic order well above \tn , i.e. up to about 100~K, which corresponds to the frustrated nature of magnetism in the P\={3}1$m$ polymorph of \mn . Observed zero-field splitting of $\Delta \approx 18$\,GHz implies small but finite magnetic anisotropy.

\section{Experimental}

Layered trigonal (P\={3}1$m$) \mn\ was prepared by ion exchange reaction as reported elsewhere.~\cite{Nalbandyan2015} Static magnetisation was studied in static magnetic fields up to 15\,T by means of a home-built vibrating sample magnetometer~\cite{vsm} (VSM) and in fields up to 5\,T in a Quantum Design MPMS-XL5 SQUID magnetometer. Pulsed field studies up to 25~T were done in a home-built device utilizing magnetic pulses of about 8 ms. Specific heat has been measured in magnetic fields up to 7~T by means of a relaxation method in a Quantum Design PPMS. Transmission HF-ESR measurements were carried out using a phase-sensitive millimeter-wave vector network analyzer (MVNA) from AB Millimetr\'{e}.~\cite{Comba2015} Experiments on a \mn\ powder sample which was fixed with eicosane were performed in the frequency range from 30 to 350\,GHz in a 16\,T superconducting magnet from Oxford instruments. ZF-$\mu$SR experiments were performed at the Paul Scherrer Institute using the DOLLY instrument with a He$^{4}$-flow cryostat in the temperature range from 2 to 250~K on a pressed powder sample ($m\sim 100$~mg). A veto-logic was used to detect only the decay positrons from muons stopped in the sample. The $\mu$SR data were analyzed with the software package MUSRFIT.~\cite{Suter2012}

\section{Static magnetization and specific heat}

\begin{figure}
\includegraphics[width=1.0\columnwidth,clip] {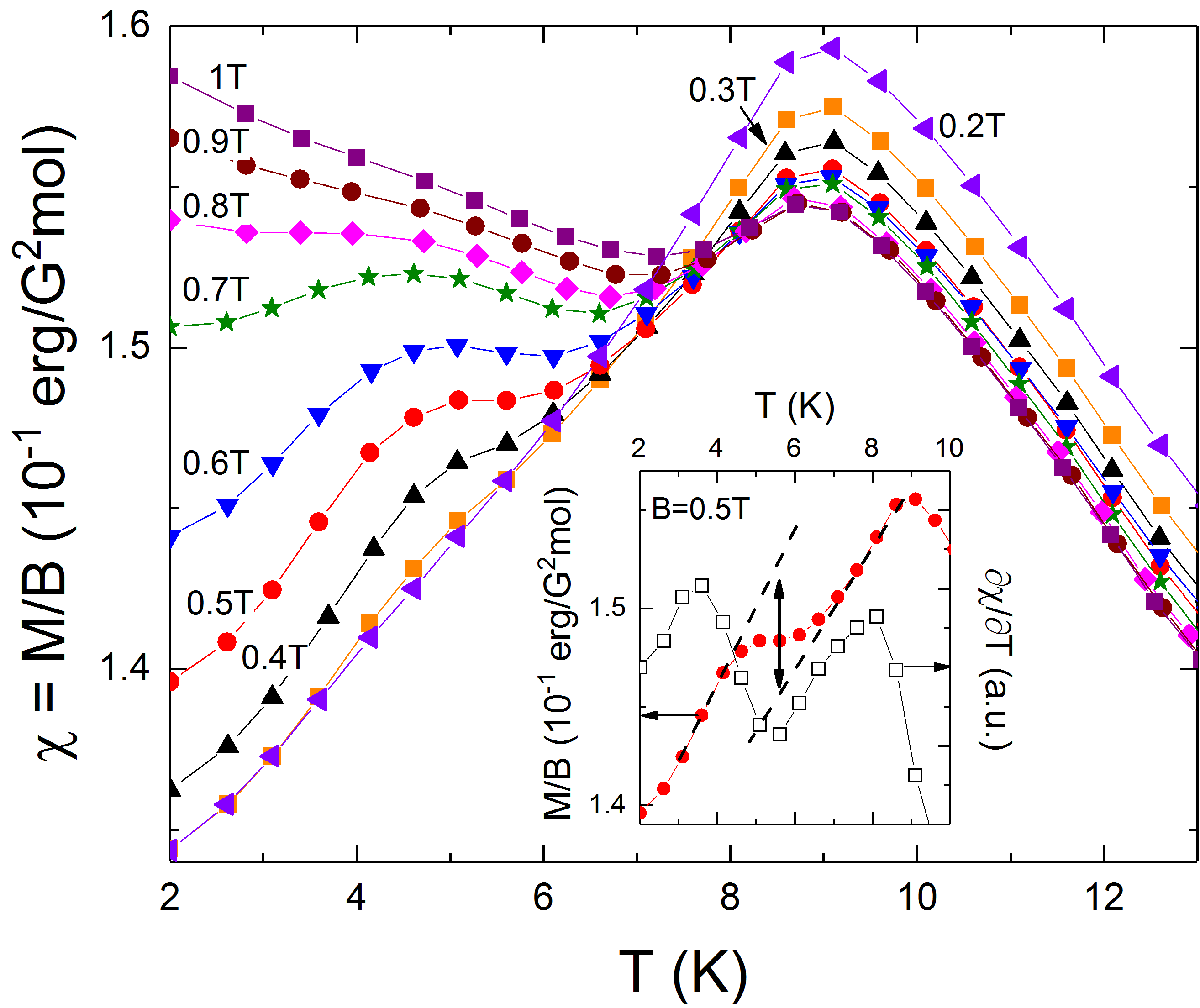}
\caption{Static magnetic susceptibility $\chi=M/B$ of \mn\ at low temperatures and at various magnetic fields $B\leq 1$~T. The inset displays $\chi$ and its derivative $\partial \chi /\partial T$, at $B=0.5$\,T, in order to highlight the anomaly at \ttwo. The dashed lines and the double-arrow indicate the jump size of the associated anomaly.}\label{MvsT}
\end{figure}

The static magnetic susceptibility $\chi=M/B$ at low temperatures shown in Fig.~\ref{MvsT} illustrates the evolution of magnetic order in \mn . While there is a broad maximum in $\chi$ vs. $T$ centered around $T=9$~K, we infer \tn\ =~8.0(4)~K from the maximum in $\partial (\chi T)/\partial T$. The negative Weiss temperature $\Theta = -20(5)$~K derived from a Curie-Weiss approximation to the high-temperature data (not shown) implies a moderate frustration parameter, i.e., $f = |\Theta |/T_{\rm N} \approx 2.6$.

While application of magnetic fields up to $B=1$~T does not significantly affect the onset of magnetic order at \tn , there are two notable features: (1) There is a general strong increase of the low-temperature susceptibility as a function of magnetic field. As will be discussed in detail below, this is associated with the spin-flop transition. (2) There is an additional anomaly, i.e., a step-like increase of the susceptibility below \tn\ at finite magnetic fields. The associated increase in the static magnetisation is illustrated in the inset of Fig.~\ref{MvsT}. E.g., at $B=0.5$\,T, there is a increase in the magnetisation of $\Delta M_{\rm N2}=6.1(2)\cdot 10^{-3}$~\mb /Mn at \ttwo\ = 5.5(2)~K. As the applied magnetic field increases, the step-like anomaly emerges, and correspondingly shifts to higher temperatures.

\begin{figure}
\includegraphics[width=1.0\columnwidth,clip] {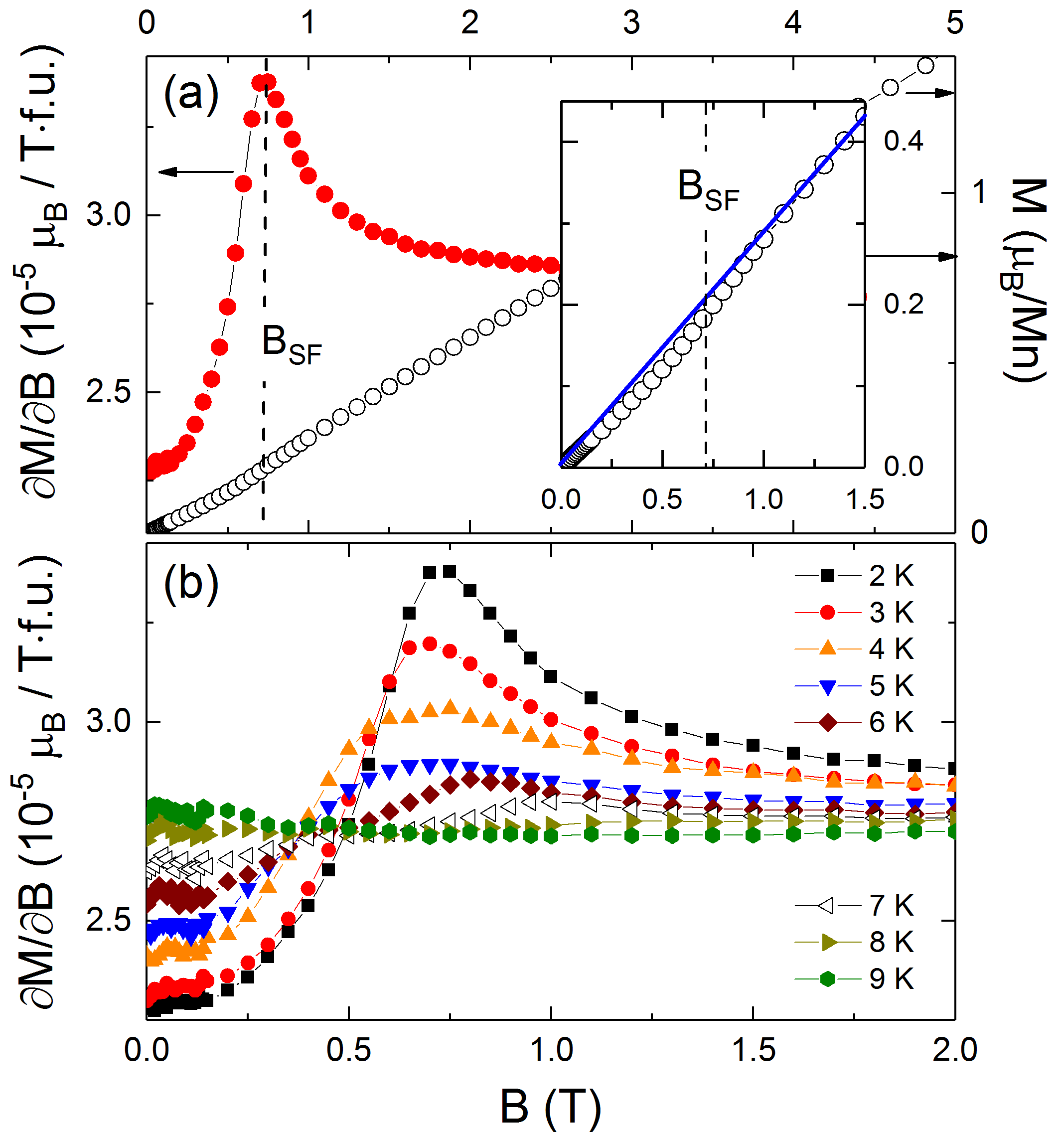}
\caption{Magnetisation $M$ and magnetic susceptibility $\partial M/\partial B$ vs. magnetic field at (a) $T=2$~K, and (b) at various temperatures up to 9~K. The inset in (a) enlarges the field range where the spin-flop occurs. The vertical dashed line indicates \bsf\ and the straight line in the inset linearly extrapolates $M$ to $B=0$~T.}\label{MvsB}
\end{figure}

\begin{figure}
\includegraphics[width=1.0\columnwidth,clip] {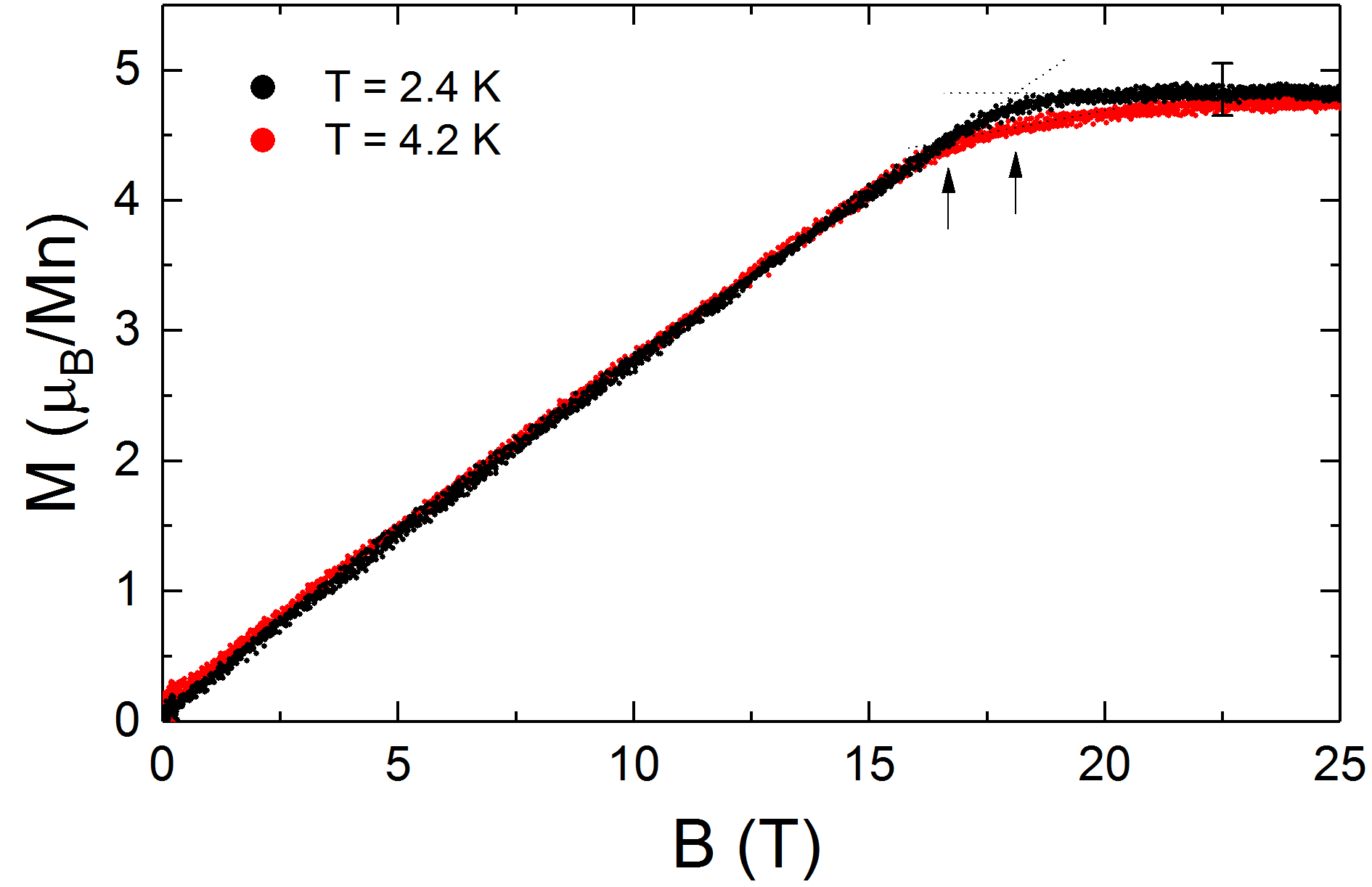}
\caption{Pulsed field magnetization obtained at $T=2.4$~K and $T=4.2$~K. Arrows indicate the saturation fields.}\label{MvsB2}
\end{figure}

The field dependence of the magnetisation shown in Fig.~\ref{MvsB} displays additional features. At $T<T_{\rm N}$, there is a small feature in $M(B)$ which is associated with a pronounced maximum in the magnetic susceptibility. This anomaly is attributed to the spin-flop transition. \rk{At 2\,K, the spin flop transition appears at \bsf\ = 0.72\,T and there is a small magnetic field hysteresis of about 0.1~T.} The corresponding jump in the magnetisation data amounts to $\Delta M_{\rm SF}\approx 0.04$~\mb /Mn. Upon heating, the anomaly at \bsf\ is suppressed and finally vanishes above \tn\ (see Fig.~\ref{MvsB}b). This scenario of spin reorientation is supported by the static susceptibility data in Fig.~\ref{MvsT} where magnetic anisotropy is evidently overcome by a magnetic field of less than 1~T. The high-field behaviour shown in Fig.~\ref{MvsB2} displays a linear field dependence of the magnetisation saturating at \bs ($T=2.4$~K) $=17.9\pm 0.1$~T. The saturation magnetisation amounts to 4.9$\pm$0.2 \mb /Mn which agrees to expected high-spin value of Mn$^{2+}$-ions.

\begin{figure}
\includegraphics[width=1.0\columnwidth,clip] {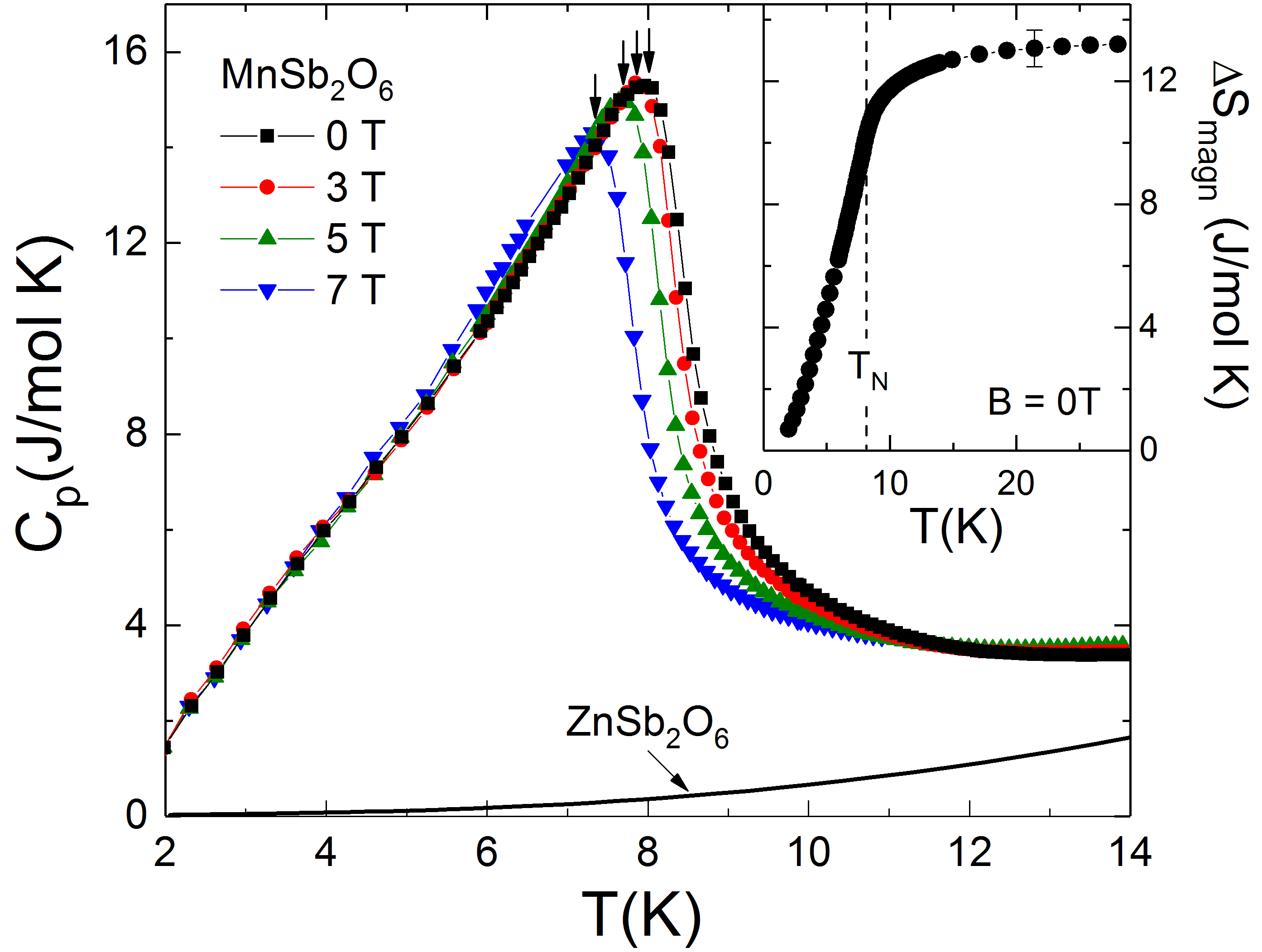}
\caption{Specific heat of \mn\ at low temperatures and at various magnetic fields and of \zn . Arrows indcate \tn ($B$). Inset: Magnetic entropy changes (see the text).}\label{cp}
\end{figure}

The effect of magnetic fields on the long-range antiferromagnetically ordered phase as well as the associated magnetic entropy changes are demonstrated by the specific heat data in Fig.~\ref{cp}. At \tn , \cp\ displays a rounded peak-like anomaly which we interpret to be remnant from a $\lambda$-shaped one. In general, the actual transition temperature might be alternatively associated with the mid of the specific heat jump at $\sim 8.6$~K.~\cite{Tari} While all our conclusions do not change with either assignment of \tn , the $\mu$SR data presented in Fig.~\ref{msr} are compatible to \tn\ $\simeq 8.0$~K. Upon application of magnetic fields, the anomaly is shifted to lower temperatures. In order estimate the magnetic entropy changes, the specific heat of the isostructural non-magnetic analogue, \zn , was subtracted from the data.~\cite{Tari} Integrating the resulting magnetic specific heat \cpm /$T$ yields the magnetic entropy changes displayed in Fig.~\ref{cp}. At \tn , about 2/3 of the total magnetic entropy $R\ln(2S+1)$ is found to be released.

In contrast to the clear anomaly at \tn , at \ttwo\ there is no clear feature in the specific heat data. \rk{The measured field dependence \ttwo ($B$) enables assessing the associated entropy or specific heat changes from the slope of the phase boundary. This procedure requires knowledge on the nature of the phase transition. Interpreting the magnetisation anomaly as a step $\Delta M$ and exploiting a Clausius-Clapeyron relation, i.e., assuming a (very weak) 1st order character yields associated entropy changes $\Delta S_{\rm N2}=-\Delta M_{\rm N2}\cdot(dT_{\rm N2}/dB)^{-1}=70(10)$~m\jmk . For a continuous transition where the anomaly is analysed as a step in $\partial M/\partial T$, the analysis yields $\Delta c_{\rm p}=-T_{\rm N2}\cdot\Delta (\partial M/\partial T)\cdot (dT_{\rm N2}/dB)^{-1}\approx 0.2$~\jmk .~\cite{klingeler2005} In both cases, the anomaly would be smeared out over $\sim 1.5$~K and cannot be resolved by the calorimetric measurements presented. In this respect we mention the ambiguous behaviour \cp ($T$) at low temperatures. The data suggest approximately linear behavior \cp\ vs. $T$ at around 5~K which exact origin is unclear. Considering the right-bending of \cp\ vs. $T$ below 3.5~K, the data are reminiscent to the presence of a hump in the temperature dependence of the specific heat in triangular spin systems such as AMn$_5$(SO$_4$)$_6$ (A = Pb,Sr) and Li$_4$FeSbO$_6$ which may be associated to frustrated nature of the spin system.~\cite{west2009,zvereva2013} Although less pronounced than in these systems, such a hump in \mn\ might overlay a possible $T^n$ behavior with $n>1$ and it prevents clear conclusions about the nature of low-energy magnetic excitations from the specific heat data.}

\section{Zero-field muon spin relaxation}

\begin{figure}
\includegraphics[width=1.0\columnwidth,clip] {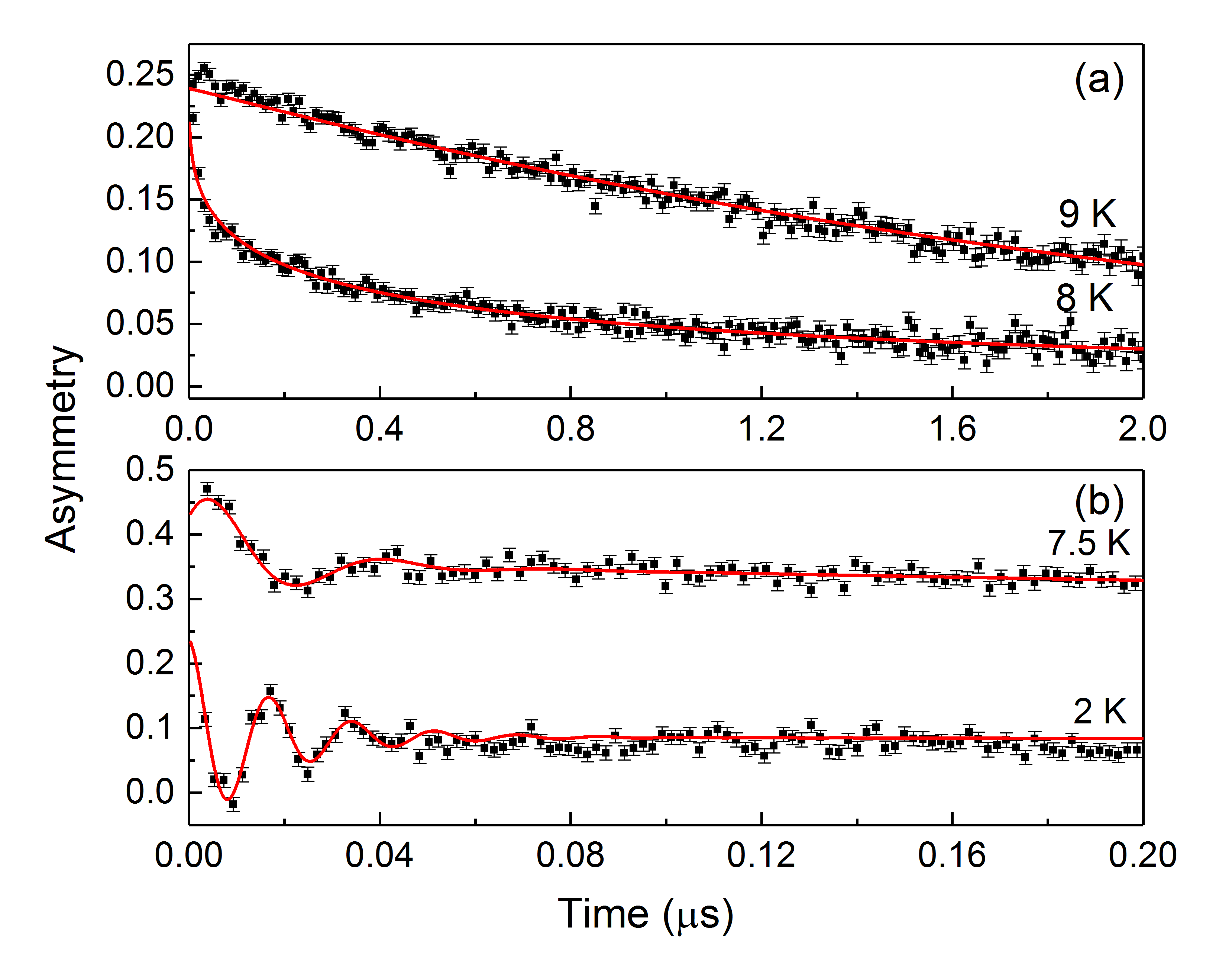}
\caption{ZF-$\mu$SR spectra of \mn\ at representative temperatures above \tn . (b) ZF-$\mu$SR spectra at early decay times below the magnetic ordering temperature \tn . The symbols and the lines represent the experimental data and the corresponding model description (for details see the main text), respectively. For clarity, the spectrum at $T=7.5$\,K in (b) is vertically shifted by 0.25, while no offset is applied to the spectra in (a). }\label{msr}
\end{figure}

In order to probe the local magnetic fields of the corresponding magnetic phase via the $\mu^+$ spin precession frequency and relaxation rate, we have performed ZF-$\mu^+$SR spectroscopy.~\cite{schenck_85} The ZF-$\mu^+$SR asymmetry spectra in Fig.~\ref{msr} confirm long-range magnetic order at low temperatures. Depending on the temperature, the spectra show two different types of behaviour. At $T\geq 9$\,K, the data exhibit only slow decay of the asymmetry signal (Fig.~\ref{msr}a). This is attributed to dipole-dipole interaction between the muon spin and the nuclear and electronic spins in \mn . In the whole temperature range 9\,K $\leq T\leq$ 270\,K (not all spectra are shown), the asymmetries are essentially identical for the first few 100~ns. In contrast, the spectra tails at $\mu$s time scales slightly increase upon heating from 9 to 20\,K. As will be discussed below, these changes of the long-time relaxation may be associated to short-range spin order. Whereas, there are no changes upon further temperature increase from 20 up to 270\,K.

\begin{figure}
\includegraphics[width=1.0\columnwidth,clip] {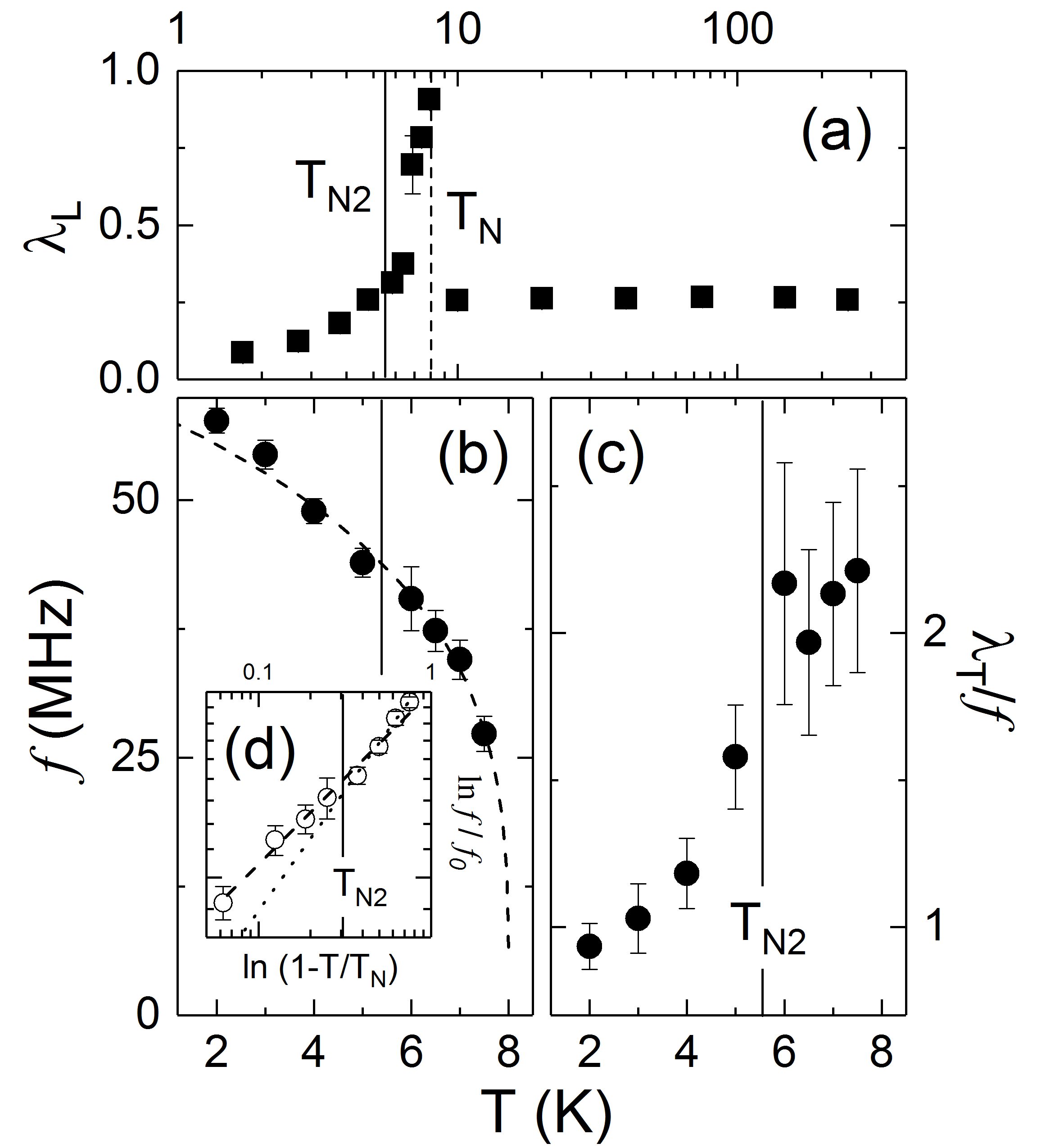}
\caption{Temperature dependence of (a) muon spin relaxation time, (b) the order parameter, i.e. the oscillation frequency {\it f}, (c) relaxation time of the oscillation divided by the frequency, $\lambda_{\rm T}/{\it f}$. \rk{(d) shows the temperature dependence of the order parameter on a double-logarithmic scale. The vertical (dashed) lines show \ttwo\ (\tn ) derived from Fig.~\ref{MvsT}. The dashed lines in (b) and (d) represent a phenomenological fit to the experimental data at \ttwo\ $\leq T\leq$ \tn\ (see the text). The dotted line in (d) is an extrapolated guide line for the data at $T<$ \ttwo .} }\label{OP}
\end{figure}

Adopting that muon-spin relaxation is dominantly due to the dynamics of electronic moments, a general exponential function $A(t) \propto \exp-(\lambda t)^{\beta '}$ was used to describe the $\mu$SR spectra above \tn . Here, $\lambda = 1/T_1$ is the generalized muon-spin relaxation rate. The exponent $\beta '$ is a measure for the homogeneity of the system. The data imply only small variation of $\beta '$ =1.15 (5) in the temperature range 9~K $\leq T\leq$ 270~K which suggests a homogeneous system. The obtained $\lambda$-values (see Fig.~\ref{OP}a) are rather temperature independent in the temperature range from 9 up to 270~K. In particular, there is no indication of local fields associated to weak ferromagnetism below 41\,K. Thus, the $\mu$SR data prove the absence of weak ferromagnetism in bulk \mn\ in contrast to Ref.~\onlinecite{Nalbandyan2015}.
%Given that the system is homogeneous, this enhancement of relaxation rate can't be justified as diluted magnetic impurity effect, rather this is associated to the short-range correlation between the electronic moments. This is reassured by the electron spin resonance experiments as discussed below.

Upon cooling below \tn , there is a clear change in the ZF-$\mu^+$SR asymmetry spectra. Above all, the asymmetry signal on the $\mu$s time scales is significantly suppressed. This becomes evident when comparing the $\mu$SR spectra at 8~K and 9~K which are displayed in Fig.~\ref{msr}a without any offset on the spectra. The asymmetry at the two temperatures is quite different in the whole time scale under study. This difference implies the presence of a fast relaxation process at $T=8$~K, implying the presence of quasi-static internal fields. The presence of long-range spin order is indeed clearly confirmed by a reasonably well defined damped oscillation behavior at $T\leq 7.5$~K (Fig.~\ref{msr}b). Upon further cooling, the damping decreases and the oscillations become more and more pronounced. The decay and the oscillation of the muon polarisation imply the evolution of local static magnetic moments in the experimental time window. In the case of a single muon site, the muon ensemble precesses about the mean local magnetic field with the Larmor frequency $2\pi {\it f}_{\mu}$ =$\gamma_{\mu}$ $|\mathbf{B}_{\rm loc}|$ where $\gamma_{\mu}$ $\sim$ 851.4~MHz/T is the gyromagnetic ratio of the muon.

As displayed in Fig.~\ref{msr}b, the damped oscillations in the $\mu$SR spectra below \tn\ are well described by means of the internal field function
\begin{equation}
  A(t) = \alpha\cdot\cos (2\pi {\it f} t + \varphi)\exp(-\lambda_{\rm T}t) + (1-\alpha)\exp(-\lambda_{\rm L}t)
\end{equation}
where $\alpha$ is the ratio of the oscillating term to the non-oscillating term, which is close to 2/3 due to the statistic average of the perpendicular orientation of the polycrystal to the direction of polarized muons. {\it f} is the Larmor frequency related to the averaged local magnetic field at the muon site, and ${\varphi}$ is the phase term. $\lambda_{\rm T}$ and $\lambda_{\rm L}$ are the muon relaxation rates reflecting the static and dynamic effects. From the frequency at $T=2$~K, i.e., $f$($T=2$~K) = 57.7~MHz, the local magnetic field at the muon site is estimated as $B_{loc}=0.43$\,T.

The oscillations are reasonably well defined and hence allow extracting the temperature dependent antiferromagnetic order parameter by means of fitting of the $\mu$SR spectra in the oscillating regime. The order parameter is presented in \rk{Fig.~\ref{OP}b and d. It shows the typical development of internal fields associated with long-range magnetic order. In a phenomenological description, the data may be described by
\begin{equation}
  f(t) = f_{0}(1-T/T_{N})^\beta .\label{eqop}
\end{equation}
The best fitting parameters describing the data at \ttwo\ $\leq T\leq$ \tn\ are $\beta = 0.28 \pm 0.02$, $f_{0}$ = 60~MHz, and \tn\ = 8~K. At \ttwo , there is a slight change in the temperature dependence of $f$ which might be associated to \ttwo\ (see Fig.~\ref{OP}d).}

The transversal relaxation rate divided by the oscillation frequency stays rather constant between 8 K and 6 K but clearly decreases for $ T \leq 5$~K (Fig.~\ref{OP}b). This change of behaviour coincides with the step-like anomaly observed in the static susceptibility measurements at \ttwo\ and the change in the temperature dependence $f(T)$. We note, however, the large error bars of $\lambda_{\rm T}$ between \tn\ and 6~K. In contrast, the muon relaxation rate $\lambda_{\rm L}$ shows no clear feature at \ttwo\ but increases as the transition temperature \tn\ is approached from below.

\section{High-Frequency Electron Spin Resonance}

\begin{figure}
\includegraphics[width=1.0\columnwidth,clip] {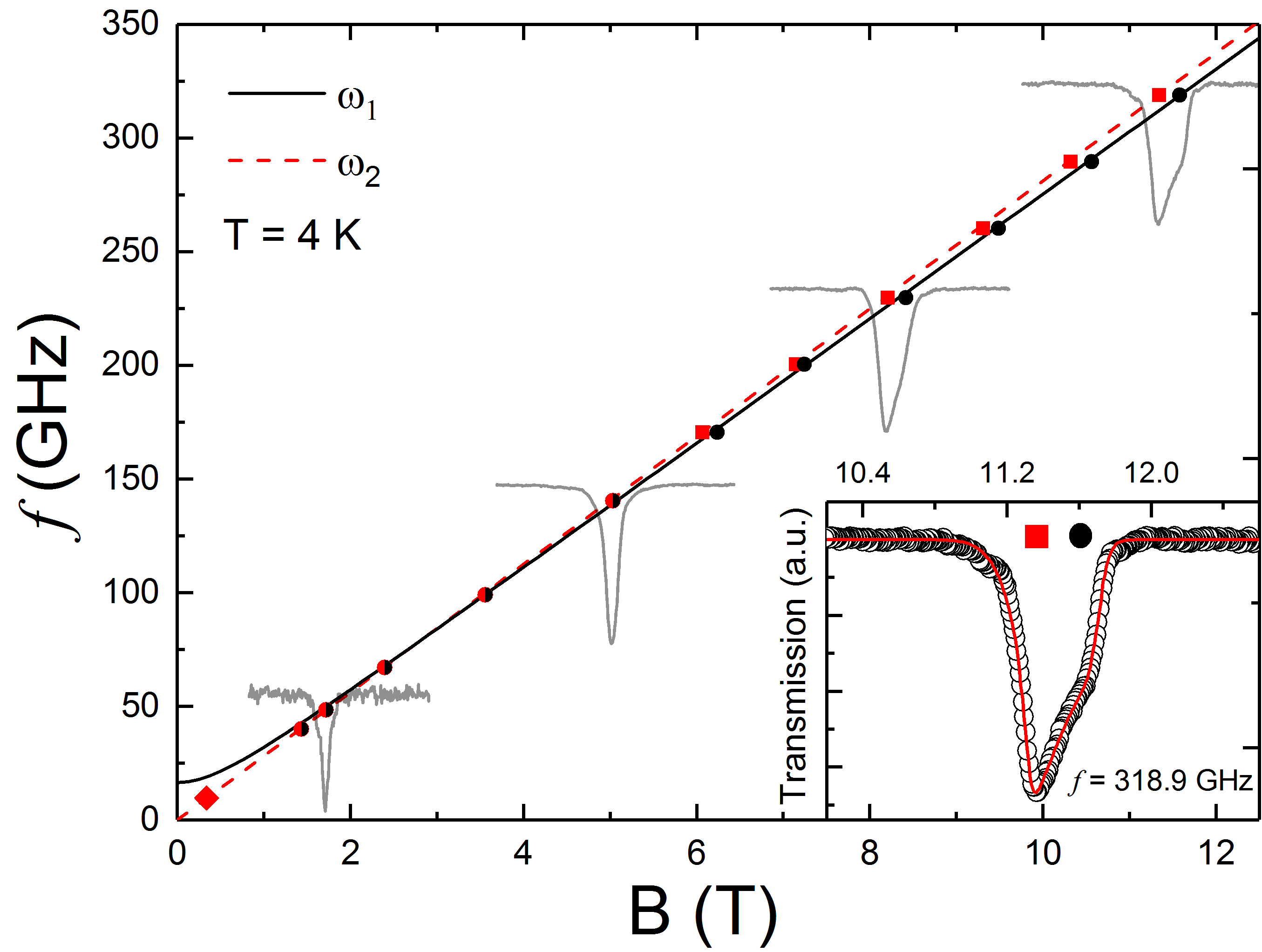}
\caption{HF-ESR absorption frequencies vs. magnetic field at $T=4$~K, together with representative spectra obtained at 50, 141, 230 and 259~GHz. The inset shows how the resonance features $\omega_{1}$ (red squares) and $\omega_{2}$ (red circles) have been obtained by fitting the resonance by means of a powder spectrum. The diamond represents the resonance field observed in previous X-band ESR measurements in Ref.~\onlinecite{Nalbandyan2015}.}\label{FESR}
\end{figure}

Further information on the magnetism in \mn\ is obtained from HF-ESR measurements. Fig.~\ref{FESR} shows selected spectra and the observed resonance fields, taken well in the long-range ordered phase at $T=4$~K, in a frequency-magnetic field plot. At high frequencies, typical transmission powder spectra with two features are observed. The more pronounced minimum feature appears at the lower magnetic field which is indicative of an easy-plane-type anisotropy.~\cite{Koo2015} Fitting the resonance features by means of powder spectra yields the respective resonance fields $\omega_{1}$ and $\omega_{2}$ shown in Fig.~\ref{FESR} which can be attributed with the magnetic fields directed parallel and perpendicular to the easy plane, respectively. At low frequencies, due to the small difference of the resonance fields we only observe single resonance features.

\begin{figure}
\includegraphics[width=1.0\columnwidth,clip] {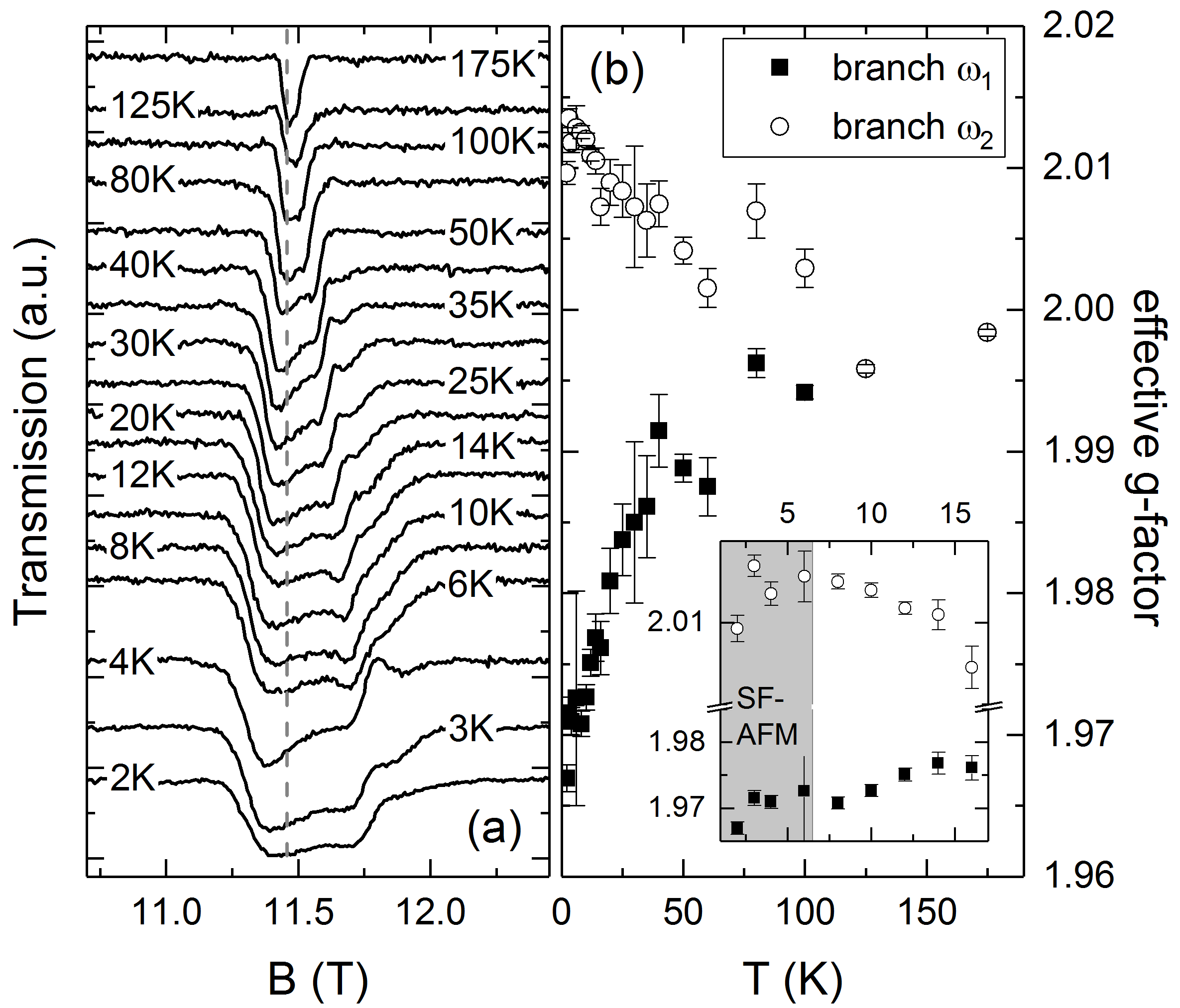}
\caption{(a) HF-ESR spectra at $f=320.2$~GHz at different temperatures between 2 and 175~K. The vertical dashed line represents $g=2$. The data are vertically shifted for clarity. (b) Temperature dependence of the effective $g$-factors associated with $\omega_{1}$ and $\omega_{2}$. The inset enlarges the low temperature region with $T\leq$ \tn (11~T) highlighted in grey. }\label{TESR}
\end{figure}

The observed resonance branches in the frequency vs. field diagram can be described by means of the antiferromagnetic resonance (AFMR) model with two sublattices in consideration of an easy-plane-type anisotropy field.~\cite{Turov} In this model, there are two AFMR resonance modes $\omega_{1}$ and $\omega_{2}$ associated with magnetic fields directed either perpendicular or parallel to the easy plane. Due to the fixation of the powder sample, we expect to observe both modes. The AFMR modes are described as following:
\begin{equation}
  \omega_{1}/\gamma =(B_{1}^2 + B_{\rm EA} ^2)^{1/2}
\end{equation}
\begin{equation}
  \omega_{2}/\gamma = B_{2}
\end{equation}
$\gamma$ is the gyromagnetic ratio. The modified magnetic fields $B_{i}=g_{i}B/2$ ($\it i$ = 1, 2) account for the anisotropy of the $g$-factor. The exchange field \be\ and the anisotropy field \ba\ are combined as $B_{\rm EA}=\sqrt{2B_{\rm E}B_{\rm A}}$. The best fit to the data yields the branches $\omega_{1}$ and $\omega_{2}$ shown in Fig.~\ref{FESR}. The best fitting parameters are $g_{1}$ = 1.97$\pm$0.01, $g_{2}$ = 2.01$\pm$0.01, and $B_{\rm EA} = 0.59\pm 0.01$~T.

While the gapless branch $\omega_{2}$ fully agrees to the resonance feature from a previous X-band ESR study~\cite{Nalbandyan2015}, our analysis suggests the presence of small but finite zero-field splitting (ZFS) of $\omega_{1}$($B=0$~T) = 18~GHz. Since no experimental data have been obtained below 30~GHz in this study due to the limitations of the experimental devices, such a small ZFS can not be unambiguously derived from the ESR data alone. The absence of any additional resonance at low frequencies down to 30~GHz however implies that the ZFS is smaller than 30~GHz. Note, that our data are not consistent with both $\omega_{1}$ and $\omega_{2}$ being gapless because in that case line splitting at 100~GHz would be large enough to be well resolved. In addition, the extracted anisotropy gap perfectly agrees to the thermodynamic spin-flop field. At $T=2$~K, the magnetisation curve demonstrates \bsf\ = 0.72~T, which corresponds to the anisotropy gap $\Delta$($T=0$~K) = $g$\mb \bsf\ $\approx $ 20~GHz. This value agrees to the best fit to the HF-ESR branches.

Figure~\ref{TESR} shows the temperature dependence of the two resonance features at $f=320.2$~GHz from 2 to 175~K. At high temperatures, the resonance is well described by a single Lorentzian with $g\sim 1.99$. Upon cooling, at temperatures not less than 100~K the resonance starts to broadening and shifting. The two resonance features can be clearly separated at $T\leq 80$~K and can be fitted with two Lorentzians. This clearly indicates the presence of local magnetic fields implying that short-range spin correlations remain well above the Neel temperatures up to at least 12\tn . As seen by the temperature dependence of the effective $g$-factors in Fig.~\ref{TESR}b, local magnetic fields continuously evolve upon cooling and eventually the typical splitting of the resonance features attributed to AFMR modes in the ordered phase is observed. Note, that no particular effects appear upon crossing the actual long-range ordering temperature \tn\ which, at the resonance field of about 11~T, amounts to $\sim 6.5$~K.

\section{Discussion}

\begin{figure}
\includegraphics[width=1.0\columnwidth,clip] {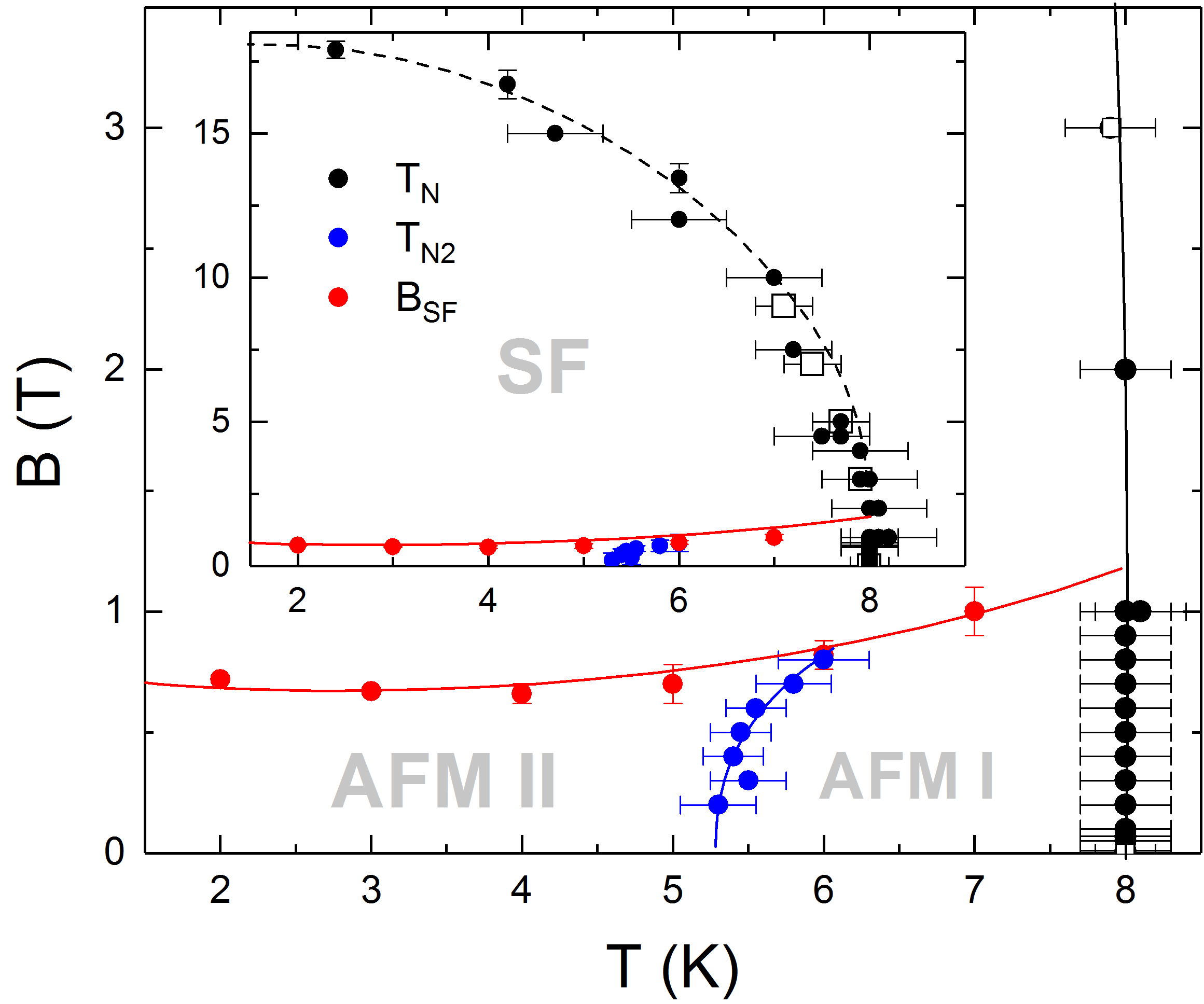}
\caption{Magnetic phase diagram of \mn\ as derived from \cp\ (open squares) , $M$ vs. $B$ and $T$ (filled circles). \tn , \ttwo , and \bsf\ indicate the onset of long-range antiferromagnetic order, the transition AFM\,I to AFM\,II, and the spin-flop transition. The inset shows the phase diagram up to high magnetic fields. All lines are guides to the eye.}\label{phd}
\end{figure}

The ZF-$\mu^+$SR-data confirm the evolution of quasi-static internal magnetic fields, i.e., of long-range AFM order in \mn\ below \tn\ $= 8.0(4)$~K. Together with the static magnetisation and specific heat data this enables constructing the magnetic phase diagram of \mn\ (see Fig.~\ref{phd}). Note, that our data clearly exclude the presence of a potential weakly ferromagnetic phase below 41~K. The phase boundary \tn ($B$) between the long-range antiferromagnetic ordered phase AFM\,I and the paramagnetic one shows the expected negative slope.

In contrast to DFT predictions~\cite{Nalbandyan2015} and different from the isostructural compound \as\ [\onlinecite{Wangbo}], magnetic order is of commensurate nature as evidenced by well-defined oscillations of the $\mu$SR asymmetry spectra. \rk{In both materials, DFT+U suggests triangular magnetic substructures in which spin exchange interactions $J_1$, $J_2$, and $J_3$ are weak and comparable in magnitude.~\cite{Wangbo,Zvereva2015} Koo and Whangbo point out that in \as\ spin exchange interactions are not the deciding factor favoring the incommensurate spin structure over a commensurate one but suggest dipole-dipole interactions being crucial.~\cite{Wangbo} The fact that our data imply commensurate spin order in the isostructural compound \mn\ where $J_1$, $J_2$, and $J_3$ from DFT+U is very similar as in \as\ confirms that even small effects may decide the actual spin ground state. We conclude that the observed effective anisotropy of 18~GHz should be considered in this respect.} \rk{In addition, our analysis of the ESR and magnetisation data suggests that the spin structure in the SF phase involves not more than two magnetic sublattices. AFMR models with three or more sublattices would imply an additional ZFS of the $\omega_2$ mode and/or a different intensity distribution of the AFMR spectra.}

Analysing the antiferromagnetic order parameter in terms of a phenomenological model yields the critical exponent $\beta = 0.28 \pm 0.02$. This value is in between the values for the 2D-XY ($\beta = 0.23$) and the 3D Heisenberg ($\beta = 0.36$) antiferromagnet which suggests that magnetic order in \mn\ is not of a simple two- or three-dimensional antiferromagnetic Heisenberg-type. This agrees to DFT calculations which yield intralayer interactions which are stronger than the (however finite) interlayer ones and would form 2D AFM networks if the interlayer coupling was negligible.~\cite{Nalbandyan2015} Hence, one neither would expect perfect 2D nor 3D magnetism in agreement with the observed critical behaviour. In contrast to the order parameter extracted from the $\mu$SR-data, splitting of the HF-ESR line does not reflect a clear change at \tn . If at all, line broadening and splitting which, in the paramagnetic phase, signals the presence of short-range correlations, stay somehow constant at \tn . This agrees to the fact that, at $f=320$~GHz, the resonance branches are well in the linear regime.~\cite{Werner2017}

At small magnetic fields, the phase diagram demonstrates two antiferromagnetic phases (AFM\,I and AFM\,II). Our observation of the yet unknown AFM II phase illustrates the presence of two competing antiferromagnetic phases. Thermodynamically, the presence of the phase boundary \ttwo\ is unambiguously derived from the associated jump in the magnetisation. In the dynamic response, the AFM II phase may be associated with a kink in the temperature dependence of the transverse relaxation rate. However, no significant anomaly in the order parameter is observed at \ttwo . In agreement with $\Delta M_{\rm N2}$ being positive, the slope of \ttwo ($B$) is positive, too. The spin-flop transition appears in both AFM I and AFM II at around 1~T. The nearly vanishing temperature dependence of \bsf\ implies negligible entropy changes at the spin-flop transition as compared to the magnetisation anomaly. In the spin-flop phase, we do not find evidence for a phase boundary \ttwo ($B$).~\cite{FN2}

Magnetic anisotropy appears to be relevant in this high-spin Mn$^{2+}$-system. The HF-ESR data imply a small but finite planar anisotropy showing up in the zero-field splitting of the associated AFMR mode of approximately 20~GHz. The resulting exchange-anisotropy field which is obtained by analysing the resonances by means of an easy-plane two-sublattice model yields $B_{\rm EA} = 0.59\pm 0.01$~T. Exploiting the saturation field $B_{\rm sat}=2B_{E}\approx 18$~T allows estimating the planar anisotropy field  $B_{\rm A}=0.02\pm0.01$~T which is the same as in trigonal P321-\mn\ and similar to CaMnCl$_3\cdot$H$_2$O.~\cite{Werner,Phaff} The thermodynamic spin-flop field quantitatively provides the same anisotropy gap as derived from the analysis of the AFMR data. Interestingly, the spin-flop phase seems not to exhibit two AFM states. \rk{This enables speculating about the relevance of anisotropy for stabilising the actual antiferromagnetic ground state. In the region of the phase diagram where the magnetic field overcomes the anisotropy energy and yields a spin-rotated situation, one of the $B=0$-spin configurations is destabilised, i.e., it only appears if anisotropy is dominant compared to the external field. We conclude that the magnetic field abrogates the energy difference between AFM\,I and AFM\,II and that magnetic anisotropy is crucial for selecting the actual magnetic state from two AFM phases of very similar energy. This conclusion is in-line with numerical results which suggest that very small effects are decisive for stabilising the incommensurate spin structure over a commensurate one in \as .~\cite{Wangbo}}

\section{Summary}

\mn\ is a spin frustrated system which magnetic structure exhibits antiferromagnetically frustrated triangles.~\cite{Nalbandyan2015} Although analysis of static magnetisation data in terms of a Curie-Weiss model only yields a moderate ratio $\Theta$/\tn\ $\approx 2.6$, the presence of local magnetic fields $viz.$ antiferromagnetic spin fluctuations is clearly confirmed by shift and broadening of the resonance HF-ESR resonance lines. The increase of the $\mu$SR spectra tail in the long time scale further confirms the presence of short-range magnetic correlations above \tn . HF-ESR data show that local magnetic fields persist up to at least $12\cdot $\tn . Such a wide temperature range of antiferromagnetic fluctuations agrees to the triangular arrangement of Mn$^{2+}$-ions and corresponding frustrated magnetism in \mn . Below \tn, muon asymmetry exhibits well-defined oscillations indicating a narrow distribution of the local fields which strongly suggests the commensurate nature of spin order. The antiferromagnetic order parameter implies a behaviour in between what is expected for the 2D-XY and the 3D Heisenberg model. A competing antiferromagnetic phase appearing below \ttwo\ =~5.3~K is evidenced by a step in the magnetisation and a slight kink of the relaxation rate. We conclude that small but finite anisotropy of $\Delta \approx 18$\,GHz derived is crucial for selecting the actual magnetic ground state from two AFM phases of very similar energy.

\begin{acknowledgements}
We are very grateful to V.B. Nalbandyan providing the samples. RK and EAZ acknowledge financial support by the Excellence Initiative of the German Federal Government and States. Partial support by the DFG via project KL 1824/13 is gratefully acknowledged. RS and HHK are thankful to the DFG for the financial support through  SFB  1143 for the project C02. This work was supported by the Ministry of Education and Science of the Russian Federation in the framework of Increase Competitiveness Program of NUST ''MISiS'' grant K2-2017-084; by act 211 of the Government of Russian Federation, contracts No. 02.A03.21.0004 and 02.A03.21.0011. Support by Russian Foundation for Basic Research through grants 17-52-45014, 18-02-00326 and 18-502-12022 is acknowledged.
\end{acknowledgements}


\begin{thebibliography}{10}
\bibitem{FN2} In the spin-flop phase such a feature would be, however, hardly visible if appearing below $T\sim 5$~K.
\end{thebibliography}
\end{document}